\documentclass[aps,prl,reprint,floatfix,superscriptaddress,showpacs]{revtex4-2}

\usepackage{amssymb}
\usepackage{amsmath}
\usepackage{amsfonts}
\usepackage{graphicx}

\newcommand{\FE}{\kappa}

\newcommand{\wv}{\vec{w}}
\newcommand{\phiv}{\vec{\Phi}}

\renewcommand{\vec}[1]{\boldsymbol{#1}}

\newcommand{\Dp}[1]{\partial_{#1}}
\newcommand{\DF}[2]{\frac{\delta #1}{\delta #2}}
\newcommand{\tDF}[2]{\tfrac{\delta #1}{\delta #2}}

\newcommand{\tsum}{\textstyle\sum}

\newcommand{\beq}{\begin{eqnarray}}
\newcommand{\eeq}{\end{eqnarray}}

\newcommand{\tr}{\text{Tr}}
\newcommand{\Tr}{{\tr}}

\newcommand{\half}{\tfrac{1}{2}}

\newcommand{\rcite}[1]{Ref.~\onlinecite{#1}}

\newcommand{\rcites}[1]{Refs~\onlinecite{#1}}

\newcommand{\Hc}{\text{Hc}}

\newcommand{\Hxc}{\text{Hxc}}

\newcommand{\Ts}{{\cal T}_{s}}

\newcommand{\E}{{\cal E}}
\newcommand{\Eb}{\bar{\cal E}}
\newcommand{\Tsb}{\bar{\cal T}_s}

\newcommand{\F}{{\cal F}}

\newcommand{\EHxc}{\E_{\Hxc}}

\renewcommand{\sd}{{\text{sd}}}
\newcommand{\dd}{{\text{dd}}}

\renewcommand{\vr}{\vec{r}}

\newcommand{\vrp}{\vec{r}'}

\newcommand{\iket}[1]{|#1\rangle}
\newcommand{\ibraket}[2]{\langle#1|#2\rangle}
\newcommand{\ibraketop}[3]{\langle#1|#2|#3\rangle}
\newcommand{\ibkouter}[1]{|#1\rangle\langle#1|}

\newcommand{\iout}{\ibkouter}

\newcommand{\up}{\mathord{\uparrow}}
\newcommand{\down}{\mathord{\downarrow}}

\newcommand{\nh}{\hat{n}}
\newcommand{\Th}{\hat{T}}
\renewcommand{\th}{\hat{t}}
\newcommand{\Wh}{\hat{W}}
\newcommand{\vh}{\hat{v}}

\newcommand{\Hh}{\hat{H}}

\newcommand{\Gammah}{\hat{\Gamma}}

\usepackage[normalem]{ulem}
\usepackage{color}
\usepackage{xcolor}

\definecolor{Mygrey}{gray}{0.80}
\definecolor{lteal}{rgb}{0.10,0.60,0.70}
\definecolor{dkred}{rgb}{0.80,0.10,0.00}
\definecolor{dkmagenta}{rgb}{0.70,0.00,0.70}
\newcommand{\comment}[1]{}

\newcommand*{\LightComments}{}%

\ifdefined\NoComments

\else
 \ifdefined\LightComments

 \else

 \fi
\fi

\begin{document}

\title{Stationary conditions for excited states:
the surprising impact of density-driven correlations}

\author{Tim Gould}
	\affiliation{Qld Micro- and Nanotechnology Centre, %
  Griffith University, Nathan, Qld 4111, Australia}%
\email{t.gould@griffith.edu.au}

\begin{abstract}
Typical density functional theory (DFT) and approximations thereto solve the many-electron ground state problem by working from a numerically efficient non-interacting Kohn-Sham reference system; and benefit from useful minimization conditions that allow iteration (i.e. self-consistency) to the optimal energy and density.
Ensembles of ground and excited states can also benefit from similar minimization conditions [{\em Phys. Rev. A} XXXX (2024)].
This work reveals that individual excited states also have state-specific stationary conditions, that can be deduced from the ensemble solution and apply to DFT and its interacting potential functional theory (PFT) counterpart.
However, the state-specific stationary condition for the non-interacting Kohn-Sham PFT is revealed to be more complicated than the ground state problem, due in part to a contribution from density-driven correlations
[{\em Phys. Rev. Lett.} {\bf 123}, 016401 (2019); {\bf 124}, 243001 (2020); {\bf 125}, 233001 (2020)]
that are neglected in ``$\Delta$SCF'' approaches.
Some implications for self-consistency in exact theory and approximations are discussed.
\end{abstract}

%%%%%%%%%%%\pacs{31.15.ec,31.15.ep,03.65.Yz}
\maketitle

In any given year, tens of thousands of papers will report density functional theory~\cite{HohenbergKohn,KohnSham} (DFT) results for chemical or material systems of interest to their authors.
These papers are almost exclusively enabled by numerical implementations of self-consistent field (SCF) theory applied to Kohn-Sham~\cite{KohnSham} (KS) DFT with density functional approximations (DFAs, e.g. B3LYP\cite{DFA:B3LYP} or PBE\cite{DFA:pbepbe}) for the Hartree, exchange and correlation (Hxc) energy term, $E_{\Hxc}$.
These implementations yield ground state energies and key properties therefrom.

Although often invisible to users, the ability to iteratively obtain an SCF solution is a vitally important aspect of KS DFT.
Specifically, the DFT energy is minimized for external potential $v$ via an effective potential, $v_s=v+v_{\Hxc}$, for which $v_{\Hxc}=\tDF{E_{\Hxc}}{n}$ is the functional derivative of the Hxc energy.
This result reflects fundamental minimization conditions enabled by the mathematical structure of exact theory and DFAs.~\cite{KohnSham,Levy1982,Yang2004-PFT}

Given their excellent performance on ground states, it is not surprising that there is interest in adapting
DFAs to excited states.
``$\Delta$SCF'' approaches (involving variational minima from ground and excited states) were first justified by G{\"o}rling~\cite{Goerling1999}, who provided the foundations for rigorous excited state DFT via extended density functionals.
Less restrictive but more specialized treatments have since been developed for nuclear potentials of form $-Z/r$~\cite{Ayers2012,Nagy2024}.
Very recently, Yang and Ayers~\cite{Yang2024-DSCF} further generalized existence conditions within potential functional theory (PFT)~\cite{Yang2004-PFT}.
A different perspective is provided by ensemble density functional theory (EDFT) for excited states~\cite{Theophilou1979,GOK-1,GOK-2}, which provides a framework for first-principles analysis of excited state problems~\cite{Gould2017-Limits,Gould2019-DD,Fromager2020-DD,Gould2020-FDT,Gould2023-ESCE}.
Recent advances in EDFT have led to first-principles-based excited state/ensemble DFAs (EDFAs)~\cite{Gould2020-Molecules,Gould2021-DoubleX,Gould2022-HL,Gould2024-GX24}.

Given that EDFAs can be constructed, an important related question is whether or not excited states (via exact theory or EDFAs) can be solved self-consistently in a safe and reliable fashion.
Various recent works -- notably from the Levi and Head-Gordon groups~\cite{Levi2020,Ivanov2021,Hait2021} -- have introduced practical orbital-based methods for finding stationary solutions that yield useful excited state energies within a DFT framework.
Giarrusso and Loos~\cite{Giarrusso2023} recently showed that stationary conditions are obeyed \emph{exactly} by excited states of a two-site Hubbard model.
Do more general stationary principles exist?

To answer this question, this letter will first summarize generalized stationary conditions for excited state ensembles, derived in an accompanying paper~\cite{Gould2024-SS-PRA}; and then extend them to individual excited states.
It will thereby establish general stationary conditions that are obyed by both ground- and excited state energies within DFT and PFT frameworks.
However, it will also reveal that the nature of excited state functionals means that the self-consistent equations have complications that are not an issue in ground states.
Some implications and a practical solution (orbital optimization) will then be discussed.
Finally, some conclusions will be drawn.

{\em Setting the stage:}
Pure state DFT is based on a fundamental variational principle, $E[v]=\min_{\Psi\to N}\ibraketop{\Psi}{\Hh[v]}{\Psi}$,  involving minimization over $N$-electron Fermionic wave functions on a Hamiltonian $\Hh[v]=\Th+\Wh+\vh$ to find the \emph{ground state} energy.
Here and throughout, $\Th$ is the many-body kinetic energy operator, $\Wh$ is the electron-electron interaction operator and $\vh=\int \nh(\vr) v(\vr)d\vr\equiv \nh\star v \equiv (\nh,v)$ the operator for external potential, $v$.
The formalism is restricted to ground states.

KS DFT invokes unique relationships between densities and potentials to work from a density, $n$, of non-interacting electronis in a fictitious potential $v_s$.
The kinetic energy of non-interacting electrons, $T_s[n]$, is then used together with an Hxc functional, $E_{\Hxc}[n]$, to capture missing many-body physics, usually via effective and low-cost DFAs.

The ensemble DFT (EDFT) of Theophilou~\cite{Theophilou1979} and Gross, Oliveira and Kohn~\cite{GOK-1,GOK-2} (TGOK) extends DFT to include excited states.
Functionally, TGOK EDFT (and most related proofs) work in much the same way as the ground state problem, except that expectation values on wave functions, $O_{\Psi}=\ibraketop{\Psi}{\hat{O}}{\Psi}$, are replaced by by ensemble expectation values, $O_{\Gammah}=\tr[\Gammah\hat{O}]$.
Thus, the ensemble counterpart of $E[v]$ is,
$\E^{\wv}[v]=\inf_{\Gammah^{\wv}}\Tr[\Gammah^{\wv}\Hh[v]]$,
where minimization is carried out over ensembles, $\Gammah^{\wv}=\sum_{\FE}w_{\FE}\iout{\Psi_{\FE}}$, formed on sets of orthogonal $N$-electron wavefunctions, $\ibraket{\Psi_{\FE}}{\Psi_{\FE'}}=\delta_{\FE\FE'}$,
defined by a set of weights $w_{\FE}\geq 0$ obeying $\sum_{\FE}w_{\FE}=1$.

By exploiting orthogonality, the TGOK variational problem yields a mixture of ground and excited state energies. Most importantly,
\begin{align}
\E^{\wv}[v]:=&\sum_{\FE}w_{\FE}E_{\FE}[v]
%=\Tr[\Gammah^{\wv}[v]\Hh[v]]
\leq  \Tr[\Gammah_{\{\Psi\}}^{\wv}\Hh[v]]
\label{eqn:Ew_bound}
\end{align}
where the right hand side involves any ensemble, $\Gammah^{\wv}_{\{\Psi\}}$,  consistent with the given weights, i.e. $\Gammah^{\wv}_{\{\Psi\}}=\sum_{\FE}w_{\FE}\iout{\Psi_{\FE}}$ for any orthonormal set of wavefunctions, $\{\Psi\}$.
The left hand side involves the eigenvalues, $E_{\FE}[v]$, of $\Hh[v]=\Th+\Wh+\vh$ and the set of weights are assigned an order (implicit throughout this work) that pairs the largest weights, $w_{\FE}$, with the smallest energies, $E_{\FE}$, i.e. $w_{\FE}<w_{\FE'}$ for $E_{\FE}>E_{\FE'}$.
Calligraphic letters and/or $^{\wv}$ superscripts indicate ensemble quantities.

EDFT also has a KS framework, with the KS kinetic energy functional~\footnote{Up to any $v_s$-representability issues -- see Sec.~V~A of \rcite{Gould2024-SS-PRA})},
$\Ts^{\wv}[n]:=\min_{\Gammah^{\wv}\to n}\tr[\Gammah^{\wv}\Th]
\equiv \min_{\Gammah_s^{\wv}\to n}\tr[\Gammah_s^{\wv}\Th]$,
found via KS ensembles, $\Gammah_s^{\wv}=\sum_{\FE}w_{\FE}\iout{\Phi_{s,\FE}}$ (indicated by subscript $_s$) consistent with non-interacting (orbital-based) Hamiltonians, $\Hh_s[v_s]=\Th + \vh_s$.
Thus, the ensemble energy functional,
\begin{align}
\E^{\wv}[v]=&\min_n \big\{ \Ts^{\wv}[n] + \EHxc^{\wv}[n] + (n,v) \big\}\;,
\end{align}
may be obtained via non-interacting (orbital) solutions, as for the ground state problem.
The universal density functional, $\F^{\wv}[n]:=\min_{\Gammah^{\wv}\to n}\tr[\Gammah^{\wv}(\Th+\Wh)]$, is used to define the Hxc energy, $\EHxc^{\wv}[n]:=\F^{\wv}[n] - \Ts^{\wv}[n]$.

The structure of $\EHxc^{\wv}$ has been explored in detail~\cite{Gould2017-Limits,Gould2019-DD,Fromager2020-DD,Gould2020-FDT,Gould2023-ESCE}, to reveal complexities that are not present in the ground state problem.
Notably:
i) the KS states are not single Slater determinants but are rather zero order renormalized orbital (ZORO) states that can reflect symmetries via finite combinations of Slater determinants; ii) there is a novel type of ``density-driven'' (dd) correlation energy that is zero in ground states~\cite{Gould2019-DD,Fromager2020-DD,Gould2020-FDT} but that contributes \emph{in addition to} the state-driven (sd) correlation energies that are analogous to (and thus allow straightforward adaptation of) ground state correlations.
Both complexities are related to the fact that the ensemble density may be written as,
\begin{align}
n^{\wv}[v]=&\sum_{\FE}w_{\FE}n_{\FE}[v]\;,
&
n_s^{\wv}[v_s]=&\sum_{\FE}w_{\FE}n_{s,\FE}[v_s]\;,
\label{eqn:nw_ens}
\end{align}
in terms of state-specific interacting densities, $n_{\FE}[v]=\ibraketop{\Psi_{\FE}[v]}{\nh}{\Psi_{\FE}[v]}$, or state-specific non-interacting densities, $n_{s,\FE}=\ibraketop{\Phi_{s,\FE}[v_s]}{\nh}{\Phi_{s,\FE}[v_s]}$.
Details shall be expanded upon as needed.

Finally, it follows from exactness that DFT corresponds to the special case $w_0=1$ and $w_{\FE>0}=0$ in $\E^{\wv}$.
Furthermore, $\Delta$SCF excited states must also be appropriate limits of ensembles.
Sometimes these pure excited states are directly accessible by the ensemble KS formalism because the TGOK variational principle is a sufficient but not necessary condition for minimization and $v\leftrightarrow n$ maps.
E.g. some excited states are obtainable via symmetry-constrained minimization~\cite{Gould2020-SP}.

{\em Stationary conditions for ensembles:}
A companion work~\cite{Gould2024-SS-PRA} has extended key variational principles of DFT to ensembles of ground and excited states; and introduced ensemble potential functional theory (EPFT) and its key variational principles.
EPFT invokes ensembles $\Gammah^{\wv}[v]=\sum_{\FE}w_{\FE}\iout{\Psi_{\FE}[v]}$ and $\Gammah_s^{\wv}[v_s]=\sum_{\FE}w_{\FE}\iout{\Phi_{s,\FE}[v_s]}$ that are respectively defined via interacting eigenstates, $\iket{\Psi_{\FE}[v]}$, of $\Hh[v]$ and non-interacting ZORO eigenstates, $\iket{\Phi_{s,\FE}[v_s]}$, of $\Hh_s[v_s]=\Th + \vh_s$.
Due to its focus on potentials, EPFT represents a closer fit to how calculations are carried out in practice and also resolves some difficulties of EDFT (see Sec.~IV of \rcite{Gould2024-SS-PRA}).

In practical terms, the goal of EDFT is to find the density that minimizes the bifunctional, $E^{\wv}[n,v^*]=\F^{\wv}[n]+(n,v^*)$; and the goal of EPFT is to find potentials $v$ and $v_s$ that minimize bifunctionals,
\begin{align}
&\Eb^{\wv}[v,v^*]=\bar{\F}^{\wv}[v] + (n^{\wv}[v],v^*)=\tr[\Gammah^{\wv}[v]\Hh[v^*]]\;,
\label{eqn:Eb_v}
\\
&\Eb_{(s)}^{\wv}[v_s,v^*]=\Tsb^{\wv}[v_s] + \EHxc[n_s^{\wv}[v_s]] + (n_s^{\wv}[v_s],v^*)\;,
\label{eqn:Eb_vs}
\end{align}
via (as required) ensemble densities $n^{\wv}[v] =\Tr[\Gammah^{\wv}[v]\nh]$ and $n_s^{\wv}[v_s] =\Tr[\Gammah_s^{\wv}[v_s]\nh]$ [see also Eq.~\eqref{eqn:nw_ens}] from interacting and ZORO states, respectively; using the universal potential functional, $\bar{\F}^{\wv}[v]=\tr[\Gammah^{\wv}[v](\Th+\Wh)]=\F^{\wv}[n^{\wv}[v]]$ and kinetic energy potential functional, $\Tsb^{\wv}[v_s]=\tr[\Gammah_s^{\wv}[v_s]\Th]=\Ts^{\wv}[n^{\wv}[v_s]]$.

The key results of \rcite{Gould2024-SS-PRA} are summarised as follows:
1) the universal and kinetic energy functionals obey,
\begin{align}
\tDF{\F^{\wv}[n]}{n}\big|_{n=n^{\wv}[v]}=&-v\;,
&
\tDF{\Ts^{\wv}[n]}{n}\big|_{n=n_s^{\wv}[v_s]}=&-v_s\;.
\end{align}
2) the potential bifunctionals [Eqs~\eqref{eqn:Eb_v} and \eqref{eqn:Eb_vs}] obey,
\begin{align}
\tDF{\Eb^{\wv}[v,v^*]}{v}=&-\chi^{\wv}[v]\star(v-v^*)\;,
\label{eqn:EnsVar_v}
\\
\tDF{\Eb_{(s)}^{\wv}[v_s,v^*]}{v_s}=&-\chi_s^{\wv}[v_s]\star(v_s-v^*-v_{\Hxc}^{\wv}[n_s^{\wv}[v_s]])\;,
\label{eqn:EnsVar_vs}
\end{align}
and are minimized when $v^*=v$ or $v_s^*=v^* + v_{\Hxc}^{\wv}[n_s^{\wv}[v_s^*]]$ (up to a constant) for the Hxc potential, $v_{\Hxc}^{\wv}[n]:=\tDF{\EHxc^{\wv}[n]}{n}$.
The mimization condition invokes the non-positive ensemble density response functions,
\begin{align}
\chi^{\wv}[v]:=&\tDF{n^{\wv}[v]}{v}
%=\tsum_{\FE}w_{\FE}\tDF{n_{\FE}[v]}{v}
= \tsum_{\FE}w_{\FE}\chi_{\FE}[v]\;,
\label{eqn:chi_v}
\\
\chi_s^{\wv}[v_s]:=&\tDF{n_s^{\wv}[v_s]}{v_s}
%=\tsum_{\FE}w_{\FE}\tDF{n_{s,\FE}[v_s]}{v_s}
= \tsum_{\FE}w_{\FE}\chi_{s,\FE}[v_s]\;.
\label{eqn:chis_vs}
\end{align}
where $\chi_{\FE}:=\tDF{n_{\FE}[v]}{v}$ and $\chi_{s,\FE}:=\tDF{n_{s,\FE}[v_s]}{v_s}$.

{\em Stationary principles for general excited states}:
The primary goal of the present work is to extend the stationary principles from ensembles to individual states.
The key enabling step is to recognise that Eq.~\eqref{eqn:Ew_bound}, and all EDFT and EPFT results that follow from it, \emph{is not restricted to weights that sum to one} -- non-negativity is the only requirement.
Although non-intuitive from a physical perspective, this result is straightfoward to justify mathematically by recognising that $W\E^{\wv}\leq W\tr[\Gammah_{\{\Psi\}}^{\wv}\Hh[v]]$ for $W>0$, and thus $\sum_{\FE} (Ww_{\FE})E_{\FE}[v]\leq \sum_{\FE} (Ww_{\FE})\ibraketop{\Psi_{\FE}}{\Hh[v]}{\Psi_{\FE}}$.
It follows that \eqref{eqn:Ew_bound} also holds for $\wv'=\{ Ww_{\FE}\}$ which sums to $W$, provided the density $n\to Wn$ and particle number $N\to WN$ are similarly scaled.
Other ensemble results follow directly; and Eqs~\eqref{eqn:Eb_v}--\eqref{eqn:chis_vs} are all well-defined functionals for \emph{any} ordered set of weights.

The second enabling step is to recognise that one may therefore take weight derivatives~\cite{Deur2019,Fromager2020-DD} of expressions to derive state-resolved counterparts, so long as the functional dependence is via a potential.
Specifically,
\begin{align}
E_{X,\FE}[v]:=\Dp{w_{\FE}}\E_X^{\wv}[v]
\equiv \lim_{\delta \to 0^+}\big( \E_X^{\wv+\delta \vec{e}_{\FE}}[v] - \E_X^{\wv}[v] \big)
\end{align}
is defined for \emph{any} ensemble functional, $\E^{\wv}_X$, so long as the weights $\wv$ obey $0<w_{\FE}<w_{\FE-1}$ -- note, $\vec{e}_{\FE}$ indicates a change made only to a single weight, $w_{\FE}$.
Degenerate energy levels may require more careful analysis but are accessisble via a similar treatment.

The first major result of this work involves taking $\Dp{w_{\FE}}$ of Eqs~\eqref{eqn:EnsVar_v} and \eqref{eqn:chi_v} to reveal that,
\begin{align}
\DF{\bar{E}_{\FE}[v,v^*]}{v}=& -\chi_{\FE}\star(v-v^*)\;.
\label{eqn:SRVar_v}
\end{align}
is stationary for \emph{all} states when $v=v^*$; for the state-specific bifunctional, $E_{\FE}[v,v^*]=\ibraketop{\Psi_{\FE}[v]}{\Hh[v^*]}{\Psi_{\FE}[v]}$.
Applying similar reasoning to the relationship $\DF{\E^{\wv}[v]}{v}=n^{\wv}[v]$ [eq.~(31) of \rcite{Gould2024-SS-PRA}] leads to $\DF{E_{\FE}[v]}{v}=n_{\FE}[v]$ and proves the relationship recently reported by Giarrusso and Loos~\cite{Giarrusso2023}.
It follows that the energy of the $\FE$th eigenstate (or its level) is stationary with respect to small changes in the potential around the self-consistent value, i.e. for $v\to v^*$.
Note, however, that it is a minimum only if $\chi_{\FE}[v]$ is non-positive, which is not the case for every state.

What about the KS system?
Taking $\Dp{w_{\FE}}$ of Eq.~\eqref{eqn:EnsVar_vs} leads to major complications.
Both $\chi_s^{\wv}$ and $v_{\Hxc}^{\wv}$ have a weight-dependence, unlike only $\chi^{\wv}$ in the interacting case.
Functional chain rules thus yield the state-resolved KS variational condition,
\begin{align}
\tDF{\bar{E}_{(s),\FE}[v_s,v^*]}{v_s}=&\chi_{s,\FE}\star(v^* + v_{\Hxc}^{\wv} - v_s)
+ \chi_s^{\wv}\star v_{\Hxc,\FE}^{\wv}
\label{eqn:SRVar_vs_bas}
\end{align}
where $v_{\Hxc}^{\wv}[n_s^{\wv}[v_s]]$ is the same Hxc potential used in \eqref{eqn:EnsVar_vs} whereas,
\begin{align}
v_{\Hxc,\FE}^{\wv}[n_s^{\wv}[v_s]]:=&\Dp{w_{\FE}}v_{\Hxc}^{\wv}[n_s^{\wv}[v_s]]
\end{align}
is its weight-deriviative contribution.
That is, each state has a variational condition that depends on its own properties \emph{and} those of the whole ensemble.

{\em Understanding eq.~\eqref{eqn:SRVar_vs_bas}}:
Where does the extra term come from?
As we shall proceed to show, it is a manifestation of transition physics that contribute to the Hartree-like term~\cite{Gould2020-FDT} and dd correlation term~\cite{Gould2019-DD,Fromager2020-DD}.
The first step is to use the fluctuation-dissipation theorem (FDT) and adiabatic connection formula (ACF) to write the Hxc energy exactly~\cite{Gould2020-FDT} as $\EHxc^{\wv}:=\sum_{\FE}w_{\FE}E_{\Hxc,\FE}$ where~\footnote{The equation is the sum of Eqs~(11), (12) and (14) in \rcite{Gould2020-FDT} written to separate the cases $\FE=\FE'$ from $\FE\neq \FE'$.},
\begin{align}
E_{\Hxc,\FE}:=&E^{\sd}_{\Hxc,\FE}[\{v^{\wv,\lambda}\}] + 2\sum_{\FE'<\FE} E_{\Hc}^{\dd}[\{v^{\wv,\lambda}\}]\;,
\label{eqn:EHxc_FDT}
\\
E^{\sd}_{\Hxc,\FE}:=&\int_0^{\lambda} \big\{ J[n_{\FE}[v^{\wv,\lambda}]] + \tr_{\omega}[\chi_{\FE}[v^{\wv,\lambda}]] \big\} d\lambda\;,
\label{eqn:EHxc_sd}
\\
E^{\dd}_{\Hc,\FE\FE'}:=&\int_0^1 J[n_{\FE\FE'}[v^{\wv,\lambda}]] d\lambda\;,
\label{eqn:EHc_dd}
\end{align}
for $J[n]:=\half\int n(\vr)n^*(\vrp) \tfrac{d\vr d\vrp}{|\vr-\vrp|}$ and $\tr_{\omega}[\chi_{\FE}]:=\half\int\{ \int_0^{\infty}\Im\chi_{\FE}(\vr,\vrp,\omega)\tfrac{d\omega}{\pi} - \delta(\vr-\vrp)n_{\FE}\} \tfrac{d\vr d\vrp}{|\vr-\vrp|}$.
Here, the densities, $n_{\FE}[v^{\wv,\lambda}]$, transition densities, $n_{\FE\FE'}[v^{\wv,\lambda}]$ and frequency-dependent response functions $\chi_{\FE}[v^{\wv,\lambda}]$ are obtained from eigenstates of the adiabatically connected~\cite{Harris1974,Langreth1975,Nagy1995,Gould2023-ESCE} Hamiltonian $\Hh^{\lambda}=\Th + \lambda\Wh + \vh^{\wv,\lambda}$.
The potential, $v^{\wv,\lambda}$, ensures that $\sum_{\FE}w_{\FE}n_{\FE}[v^{\wv,\lambda}]=n^{\wv}\forall\lambda$ and obeys $v^{\wv,0}[n]=v_s^{\wv}[n]$ and $v^{\wv,1}[n]=v$.
But, because \eqref{eqn:EHxc_sd} and \eqref{eqn:EHc_dd} depend on $v^{\wv,\lambda}$ they do not lend themselves to detailed analysis.

The next step is to follow recent work~\cite{Gould2024-GX24} that invoked the ACF, FDT and low-density limit of matter~\cite{Gould2023-ESCE} to devise a first-principles approximation that depends only on $v_s$. Specifically,
\begin{align}
E^{\sd}_{\Hxc,\FE}\approx&  E_{\Hxc}^{\text{DFA}}[n_{s,\FE}]\;,
&
E_{\Hc}^{\dd}\approx& (1-\xi)J[n_{s,\FE\FE'}]
\label{eqn:EHxc_vs_S}
\end{align}
invokes an sd-Hxc energy from $\Delta$SCF-like application of DFAs to state-specific densities $n_{s,\FE}[v_s]$,
and a dd-Hc energy that contains the dd correlations and H-like terms involving transition densities, $n_{s,\FE\FE'}[v_s]$ to lower-lying ZORO states, $\iket{\Phi_{s,\FE'}[v_s]}$.
Remarkably, the dd-Hc terms appear even in \emph{pure state} treatment of some excited states; and its inclusion in the GX24 EDFA leads to impressive excited state prediction~\cite{Gould2024-GX24}.

Eq.~\eqref{eqn:EHxc_vs_S} allows direct evaluation of the stationary condition.
Taking $\tDF{}{v_s}$ of \eqref{eqn:EHxc_FDT} leads to,
$\chi_{s,\FE}\star \bar{v}_{\Hxc,\FE} \approx \chi_{s,\FE}\star v^{\sd}_{\Hxc,\FE}[n_{s,\FE}]
+ 2\sum_{\FE'<\FE} \chi_{s,\FE\FE'}\star v^{\dd}_{\Hc}[n_{s,\FE\FE'}]$,
where $\bar{v}_{\Hxc,\FE}:=\tDF{\bar{E}_{\Hxc,\FE}}{n}$, $v^{\sd}_{\Hxc,\FE}:=\tDF{E^{\sd}_{\Hxc,\FE}}{n}$, $v^{\dd}_{\Hc}:=\tDF{E_{\Hc}^{\dd}}{n}$ and $\chi_{s,\FE\FE'}:=\tDF{n_{s,\FE\FE'}[v_s]}{v_s}$.
Taking $\tDF{}{v_s}$ of $\EHxc^{\wv}[v_s]\approx \sum_{\FE}w_{\FE}\bar{E}_{\Hxc,\FE}[v_s]$ yields $\chi_s^{\wv}\star v_{\Hxc}^{\wv} \approx \sum_{\FE} w_{\FE} \chi_{s,\FE}\star \bar{v}_{\Hxc,\FE}$; and taking $\Dp{w_{\FE}}$ yields, $\chi_{s,\FE}\star v_{\Hxc}^{\wv} + \chi_s^{\wv}\star v_{\Hxc,\FE}=\chi_{s,\FE}\star \bar{v}_{\Hxc,\FE}$.

Finally, using the above results in \eqref{eqn:SRVar_vs_bas} leads to,
\begin{align}
\DF{E_{(s),\FE}[v_s,v^*]}{v_s}=&\chi_{s,\FE}\star(v^* + \bar{v}_{\Hxc,\FE} - v_s)
\label{eqn:SRVar_vs_good}
\end{align}
which is a good stationary condition for KS potentials --  here, equality indicates that Eq.~\eqref{eqn:SRVar_vs_good} is exact for Eq.~\eqref{eqn:EHxc_vs_S}.
Note, however, that,
\begin{align}
\bar{v}_{\Hxc,\FE} = v^{\sd}_{\Hxc,\FE} + \sum_{\FE'<\FE}\chi_{s,\FE}^{-1}\star \chi_{s,\FE\FE'}\star v^{\dd}_{\Hc,\FE\FE'}\;,
\label{eqn:SRVar_vs_ugly}
\end{align}
involves $\chi_{s,\FE}^{-1}\star \chi_{s,\FE\FE'}$ which is an indirect and non-local functional of $v_s$.
This is the second major result of this work as it reveals that the stationary principle for non-interacting KS excited states has contributions from lower levels, unlike the state-specific stationary principle [Eq.~\eqref{eqn:SRVar_v}] for interacting potentials.

A mathematically similar result can also be obtained from exact theory.
First, recognise that $w_{\FE}\bar{E}_{\FE}[v,v^*]=(1-\sum_{\FE'\neq\FE}w_{\FE'}\Dp{w_{\FE'}})\Eb^{\wv}[v,v^*]$ follows directly from $\Eb^{\wv}[v,v^*]=\sum_{\FE}w_{\FE}\bar{E}_{\FE}[v,v^*]$ and the properties of partial derivatives.
Then, define $v_{\Hxc}^{\wv}[v]$ such that $n_s^{\wv}[v+v_{\Hxc}^{\wv}[v]]=n^{\wv}[v]$ and $\Eb^{\wv}[v,v^*]=\Eb_{(s)}^{\wv}[v+v_{\Hxc}^{\wv}[v],v^*]$;
and take $\Dp{w_{\FE}}$ of $w_{\FE}\bar{E}_{\FE}[v,v^*]=(1-\sum_{\FE'\neq\FE}w_{\FE'}\Dp{w_{\FE'}})\Eb_{(s)}^{\wv}[v_s^{\wv}[v],v^*]$ to obtain,
$\bar{E}_{(s),\FE}[v_s^{\wv},v^*]
=\bar{E}_{\FE}[v,v^*]
+ \sum_{\FE'\neq\FE}\frac{w_{\FE'}}{w_{\FE}}v_{\Hxc,\FE'}^{\wv}\star\tDF{\Eb_{(s)}^{\wv}[v_s^{\wv},v^*]}{v_s}$,
because both $v_{\Hxc}^{\wv}[v]$ and $\bar{E}_{(s)}^{\wv}$ depend on $w_{\FE}$.
Finally, the special ensemble weights $\vec{t}_{\FE}$ obeying $t_{\FE'\leq \FE}=1$ and $t_{\FE'>\FE}=0$ yield,
\begin{align}
\bar{E}_{(s),\FE}^{\vec{t}_{\FE}}[v_s^{\vec{t}_{\FE}},v^*]=&\bar{E}_{\FE}[v,v^*]
+ \sum_{\FE'<\FE}v_{\Hxc,\FE'}^{\vec{t}_{\FE}} \star
\tDF{\Eb_{(s)}^{\vec{t}_{\FE}}[v_s^{\vec{t}_{\FE}},v^*]}{v_s}\;,
\label{eqn:E_S_lower}
\end{align}
where $v_{\Hxc,\FE}^{\vec{t}_{\FE}}\equiv v_{\Hxc,\FE}^{\vec{t}_{\FE}}[n_s^{\vec{t}_{\FE}}[v_s]]$.
The first term of Eq.~\eqref{eqn:E_S_lower} naturally pairs with the sd-Hxc term and $\Delta$SCF approximations. The second term pairs with dd-Hc energies in Eq.~\eqref{eqn:EHxc_vs_S} and it follows that the above steps [including Eqs~\eqref{eqn:SRVar_vs_good} and \eqref{eqn:SRVar_vs_ugly}] may be replicated for the exact case.

\begin{figure}[t!]
\includegraphics[width=\linewidth]{{{FigCompare}}}
\caption{Orbital optimized singlet-singlet excitation energies [eV] from both terms of Eq.~\eqref{eqn:EHxc_vs_S} (left) and the $\Delta$SCF term only (middle); and errors (right).
Blue stars are closed shell and orange stars are radicals.
Results for single (acrolein, butadiene, cyanoformaldehyde, cyclopentadiene and tetrazine) and double excitations (BH radical, formaldehyde, glyoxal, and nitroxyl). Reference data from \rcites{Loos2018,Loos2018a,Loos2019,Gould2024-GX24}.
\label{fig:Compare}}
\end{figure}

{\em Implications}:
On reflection, it is not so surprising that the stationary condition depends on lower energy states.
After all, an excited states is only an excited state \emph{because} there are states lower in energy.
For the interacting system, the map from $v\to n$ and $n\to v$ can (in some sense) detect the ground or excited state.~\cite{Goerling1999,Nagy2024}
But for the \emph{fictitious} non-interacting KS system this is not the case as the implicit map $v_s\leftrightarrow v$ must retain information about energy levels \emph{and} densities, as manifested in Eqs~\eqref{eqn:EHxc_vs_S} or Eq.~\eqref{eqn:E_S_lower}.
It stands to reason that lower energy states can have an impact on exact functionals, approximations, and self-consistent cycles thereto.

What impact does this have in practice?
Firstly, one may retain only the sd-Hxc term in Eq.~\eqref{eqn:EHxc_vs_S}.
Then, the extra potential terms disappear and a self-consistency cycle (modified for stationary solutions) can yield the KS potential for any given state.
However, neglecting the dd-Hc term in approximations may miss important physics.

A practical alternative that retains the important physics is to use \eqref{eqn:EHxc_vs_S} in full, but work with orbital optimization to seek energies that are stationary with respect to a set of orbitals, $\phiv=\{\phi_i\}$.
That is: i) use the properties of ZORO states to write $n_{s,\FE}[\phiv]=\sum_i f_i^{\FE}|\phi_i|^2$ [per \eqref{eqn:nw_ens}] and $n_{s,\FE\FE'}[\phiv]=\sum_{i\neq j}c_{ij}^{\FE\FE'}\phi_i^*\phi_j$ where $f_i^{\FE}$ and $c_{ij}^{\FE\FE'}$ are constants that depend on the states; ii) substitute the expressions in \eqref{eqn:EHxc_vs_S}; and iii) seek $\tDF{\bar{E}_{(s)}^{\wv}[\phiv,v^*]}{\phi_i}$ or $\tDF{\bar{E}_{(s),\FE}[\phiv,v^*]}{\phi_i}=0$ subject to $\ibraket{\phi_i}{\phi_j}=\delta_{ij}$.

For ensembles, Sec.~VI of \rcite{Gould2024-SS-PRA} reveals that the minimizing set of orbitals $\phiv^{\wv,*}=\{\phi_i^*\}$ can be used to bound the EPFT energy via,
\begin{align}
\Eb^{\wv}_{(s)}[\phiv^{\wv,*},v^*]\leq \Eb^{\wv}_{(s)}[v_s^*,v^*]
\leq \Eb^{\wv}_{(s)}[v_s[\phiv^{\wv,*}],v^*]
\label{eqn:Ew_Bounds}
\end{align}
where $v_s[\phiv]$ indicates a potential for which $n_s^{\wv}[\phiv]=n_s^{\wv}[\phiv[v_s]]$ for orbitals $\phiv[v_s]$ obeying $\{\th + v_s\}\phi_i[v_s]=\epsilon_i[v_s]\phi_i[v_s]$.
The upper and lower bounds may be equalized in any finite basis set by exploiting variational freedoms in potentials.

Sometimes, orbital optimized solutions, $\phiv^*_{\FE}$,  for specific excited states can be found via minimization of the excited state energy expression, subject to constraints on occupations.
Then, provided all $\chi_{s,\FE'}^{\wv}$ for $\FE'\leq \FE$ are negative definite~\footnote{This ensures that minimization conditions apply to all possible selections of positive weights, including those that violate the TGOK conditions.} in a sufficiently large neighbourhood around stationary solutions, one obtains,
\begin{align}
\bar{E}_{(s),\FE}[\phiv^*_{\FE},v^*]
\leq \bar{E}_{(s),\FE}[\bar{v}_{s,\FE}^*,v^*]
\leq \bar{E}_{(s),\FE}[v_s[\phiv^{\wv,*}],v^*]\;,
\label{eqn:ES_Bounds}
\end{align}
where the interior expression inovkes the stationary solution, $\bar{v}_{s,\FE}^*=v^* + \bar{v}_{\Hxc,\FE}^*$ of Eq.~\eqref{eqn:SRVar_vs_good}.
The lower bound follows from variational principles and the upper bound follows from ensemble solutions minimizing the ensemble energy, not the energies of individual states.
The upper bound in a finite basis set may be replaced by $\bar{E}_{(s),\FE}[\phiv^{\wv,*},v^*]$.

Figure~\ref{fig:Compare} shows orbital optimized singlet-singlet excitation energies computed using GX24 [sd-Hxc and dd-Hc in Eq.~\eqref{eqn:EHxc_vs_S}] and conventional $\Delta$SCF [sd-Hxc only, using a pure state DFA for closed shells, and assigning $\up$ to one unpaired orbital and $\down$ to the other in radicals].
The two approaches agree closely for closed shell excitations.
But, underestimation of radical energies with $\Delta$SCF are systematically corrected by the dd-Hc transition physics, highlighting its importance.
All calculations use the same DFA~\cite{Gould2024-GX24} and orbital optimization procedure~\footnote{Using {\tt Broadway}/v1.5 code with the cc-pvdz basis set. Source code available on request.}.

{\em Conclusions:}
This work has shown [eq.~\eqref{eqn:SRVar_v}] that \emph{all} ground and excited states obey stationary conditions within PFT, for functional derivatives with respect to external potentials.
However, when KS potentials are employed as the basic variable, the problem becomes more complicated [eqs~\eqref{eqn:SRVar_vs_bas}, \eqref{eqn:SRVar_vs_good}, \eqref{eqn:SRVar_vs_ugly}].
In these cases, density-driven (dd) correlations and H-like terms give rise to complications which mean that natural choices for Kohn-Sham potentials, $v_s^{\wv}=v+\tDF{E^{\wv}_{\Hxc}}{n}$ or $v_{s,\FE}=\Dp{w_{\FE}}v_s^{\wv}$, do not yield the variational minimum of energies with respect to KS potentials.
Specifically, solutions must capture the effect of lower-lying states [eqs~\eqref{eqn:EHxc_vs_S} and \eqref{eqn:E_S_lower}] or risk underestimating excitation energies when using $\Delta$SCF terms only (Figure~\ref{fig:Compare}).

The present work also directly connects $\Delta$SCF approaches and EDFT.
Due to the technological and scientific need to understand and model optical excitations, there is an urgent need for novel excited state functionals and methods to solve them variationally.
As shown here, and demonstrated in the GX24 EDFA~\cite{Gould2024-GX24}, accurate excited state functionals may need to account for ensemble-informed and conventional DFA physics.
But, ensemble physics makes SCF much more difficult.
Orbital opimization algorithms to solve for stationary conditions should thus continue to be refined~\cite{Levi2020,Ivanov2021,Hait2021}.
Extension to ensembles that capture addition or removal of an electron~\cite{Cernatic2022,Cernatic2024-SingleX,Cernatic2024-DoubleX} should also be pursued.

\acknowledgments
{\em Acknowledgments:} 
This research was funded through an Australian Research Council (ARC)
Discover Project (DP200100033).
TG acknowledges support from an ARC Future Fellowship (FT210100663).
TG thanks Sara Giarrusso and P.F. Loos for feedback
on an early draft of this work

%\bibliography{StatCond}

\end{document}